# The Structure of Liquid Water Emerging from the Vibrational Spectroscopy: Interpretation with QED Theory

A.De Ninno[1], E.Del Giudice[2,] L.Gamberale[3], A. Congiu Castellano[4]

[1]C.R. ENEA Frascati Department UTAPRAD-DIM 00044 Italy, antonella.deninno@enea.it
[2]Retired Scientist, Milano Italy, emilio.delgiudice@mi.infn.it
[3]MOSE srl, Milano Italy , luca.gamberale@gmail.com
[4]Department of Physics, Sapienza University of Rome 00185 Italy, a.congiu@caspur.it

## Abstract

We report an analysis of the stretching peak appearing in the IR experimental spectra of liquid water. In the literature, ATR-IR spectroscopic measurements were repeatedly performed in a wide range of temperature and gave rise to a lively debate among scientists. In particular a two components model related to H-bond complexes of different strength have been proposed in order to justify the existence of two types of molecules as it appears from the spectroscopic data. At the opposite, Molecular Dynamics simulations support a multistate (continuum) system of H-bond having different strength giving rise to a (locally) tetrahedral description of liquid water. We will show that liquid water is a quantum two-level system according to the predictions of Quantum Electrodynamics (QED) and that several features (the asymmetric band profile, the existence of an isosbestic point and the modifications of the vibrational stretching band with the temperature) cannot be fully justified in the realm of a classical picture. In particular the differences of energy and entropy between the two phases are estimated from the experimental data and compared with the prediction of QED showing a remarkable agreement. The behaviour of water near hydrophilic surfaces is also discussed and several feature of the so called "Exclusion Zone" observed by several authors are evaluated according to the two level system model.

# 1 Introduction

The structure of water has been studied by a variety of techniques: Raman scattering, neutron scattering, vibrational spectroscopy. The clarification of the structure of water is considered fundamental to understand both the evolution of life and its specificity. In spite of the large number of papers published in the last twenty years in the literature, the basic model of liquid water is still highly controversial. The molecular dynamics (MD) modelling of the shapes of the $H_2O$ IR and Raman spectra could seem to support the continuum model of water [1,2,3]. According to the viewpoint of liquid-state statistical mechanics the continuum picture emerges quite naturally in case the system is at a sufficiently high temperature and low density in order to assume the mutual independence of the component molecules so that the ensemble of molecules should be described by a density matrix and not by a unique wave function. In case of "isolated" chromophores, the transition frequency of a vibrational mode will fluctuate on some characteristic time scale. In case this time rate is sufficiently slow the line shape will be simply the distribution of frequency; at the opposite, the effect of fluctuation dynamics will appear into the spectrum and the line shape will be narrower than the distribution of the frequencies. However, in case of liquid water the situation is quite more complicated because of the intermolecular and intramolecular coupling. One recently proposed solution, the Time Averaging Approximation (TAA) calculates the line shape as a distribution of time averaged frequencies where the time of averaging is a parameter obtained by fitting the experimental line shape [4]. However, these calculations are not intended to model spectra of real water because they do not include intermolecular interactions. The line shape obtained by using TAA, even including non-Condon effect and rotations is essentially Gaussian but this shape doesn't fit correctly the experimental line shape.

The analysis of the line shape variation with the temperature shows that it can be exactly decomposed in two basic spectra. This is an experimental fact and cannot be ignored, it strongly suggests the presence of two kinds of molecules in liquid water [5,6,7,8,9,10]. This point has been the object of a very lively discussion between alternative explanations about the existence of a continuum distribution of H-bonds instead of the presence of two kinds of molecules in $H_2O$. It has been suggested also that these two pictures are not mutually exclusive since each type of molecules carries with it a continuous broad distribution of H-bonds [5], however, in such a context, it is not clear the physical meaning of the isosbestic point which suggests the existence of two populations at equilibrium according to the Van't Hoff equation.

In this paper we will show that the well known spectroscopic data of liquid water, here reproduced using ATR-IR spectroscopy in order to have a set of experimental data on which perform the calculations are fully compatible with the prediction of Quantum Electrodynamics (QED). Actually it has been shown starting from first principles that, above a density threshold and below a critical temperature, a coherent state is induced in the physical system. Water represents a remarkable example of such a general principle. Such a picture accounts for the spectroscopic features of liquid water, including the quantitative estimation of the broadening of the vibrational line shape without the use of any adjustable parameter and the energy gap between the two species of molecules but is also able to explain the features of an interesting variety of liquid water reported in literature, the so called Exclusion Zone (EZ), which is liquid water detected very close to hydrophilic surfaces.

2 Experimental results

Vibrational spectroscopy is a well proven method capable to characterize water properties on the molecular level. Since a bi-univocal correspondence exists between the OH stretching frequencies and the degree of binding, it is possible to obtain structural information from mid IR spectra. We have measured the IR spectrum of the O-H stretching in liquid water in the range of temperature 298-333 K. Briefly, the IR measurements have been performed using an FTIR/410 Jasco Fourier Transform IR spectrometer (Jasco International Co. LTD, Tokyo, Japan) equipped with an Attenuated Total Reflectance (ATR) accessory (ATRPRO410-S) with a 45° single reflection ZnSe crystal plate. A resolution of 4 cm$^{-1}$ has been obtained for each spectrum, and 120 interferograms have been co-added and apodised using a triangular function. The preliminary reduction of the IR spectral data (subtraction, normalisation, ATR penetration depth compensation) and the fitting have been performed using the Jasco Spectral Analysis (Jasco International Co. LTD, Tokyo, Japan) software.

Careful inspection of the line shape does not show a unique Gaussian line, actually it is well known that the broad absorption peak from 2800 to 3800 cm$^{-1}$, attributed to the OH stretching mode, can be deconvoluted into line shapes which are usually assigned to different sets of molecules. The deconvolution in sub-bands could be done mathematically in infinitely many ways; this is a trivial consequence of the central limit theorem which states that independent random variables (in the present case the number of H-bonds) are normally distributed, i.e. tend to be distributed around a mean value. Different sub-distributions, on the base of several (up to eight) Gaussians are based on specific hypothesis on the numbers and the strength of different H-bonding configurations [11]. However the hypothesis should be physically reasonable and able to be checked experimentally. Several features coalescing into a broader one should reasonably maintain their fingerprint in the convoluted line; this fingerprint has been identified with the presence of a minimum of the second derivative of the convoluted line which corresponds to the presence of an underlying peak. Consequently the second derivative analysis of the IR spectrum in the OH stretching region is used to find the centres of the absorbance band of each component existing inside the broad absorption peak. The second derivative spectrum in Fig. 1 shows three sub bands and provides the position of each peak at (about) 3200 $cm^{-1}$, 3400 $cm^{-1}$ and 3600 $cm^{-1}$.

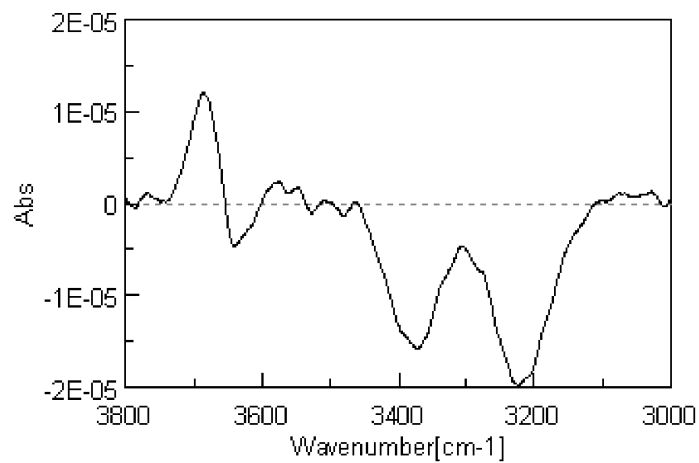

Figure 1. Second derivative spectrum of liquid water at 298 K.

The above observations introduce a univocal decomposition of the broad absorption line since are based on structural characteristic of the spectrum. Since the OH bond is sensitive to the environment, these bands have been assigned to different degrees of interactions of the concerned molecules. It is quite interesting to address the problem of the dependence of the peaks on the interactions of the involved molecules by comparing the spectroscopic features of gaseous, liquid and solid water as shown in Fig. 2 which reproduces the figure exhibited in [12]. It is apparent that in vapour, where molecules can be assumed safely to be independent, the peaks at 3200 and 3400 $cm^{-1}$, observed in condensed water are absent, whereas we can observe a number of peaks whose position is slightly higher than 3600 $cm^{-1}$. On the contrary in ice the dominant feature is the peak at 3200 $cm^{-1}$ while the other features fade away by decreasing temperature. It is worthwhile noting that in ice above 233 K it is still present an amount of included liquid water responsible for the structures observed in the peak. Since we know that component molecules in ice are highly structured we can assume that the peak at 3200 $cm^{-1}$ is produced by the sub ensemble of molecules which are mutually interacting. In the liquid, where we can observe three peaks, we should assume the simultaneous presence of a fraction of locally interacting molecules (monomers and dimers such as in the gas phase), a fraction of molecules having highly structured interactions (such as in ice) and an intermediate fraction (responsible of the peak at ~ 3400 cm$^{-1}$). We will consider this intermediate fraction further in the discussion.

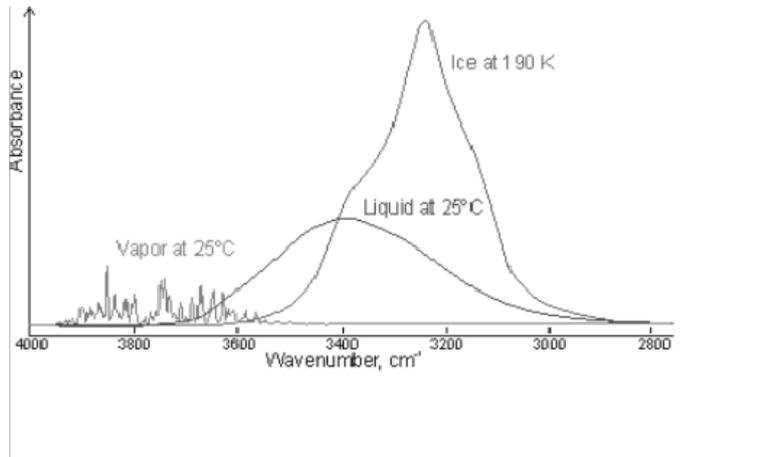

Figure 2. Comparison of the gas, liquid and solid spectra of the same amount of water. From Martin Chaplin: Water Structure and Science web page http://www.lsbu.ac.uk/water/vibrat.html

The coalescence of the peaks observed in the dilute vapour in a unique peak in liquid (Fig.2) is presumably related to the much higher density, and correspondingly to the much closer intermolecular distance, of the non-interacting molecules in the liquid. We can also underline that the apparent homogeneity of the liquid should imply that the type of interactions responsible for the presence of two well separated sub-peaks is not connected with the intermolecular distance and therefore to the binding of molecules produced by short range forces (which are presumably present in both sub-ensembles). Such an interaction should depend by a different physical agent as we will discuss in the next section. On the basis of the above experimental observations, let's assume that the spectral intensity $I(\omega, T)$ can be decoupled in two components having a different intermolecular dynamics. Let us call them fraction (1) and fraction (2), the latter being the sum of what we call intermediate and almost free molecules, both fraction have been observed to depend on temperature T:

$$I(\omega,T) = I_1(\omega,T) + I_2(\omega,T) \qquad (1)$$

It has been reported [13,14,15] that $I(\omega,T)$ shows an isosbestic point upon changing the temperature. The isosbestic point is the frequency value $\omega$, at which the spectra $I(\omega,T)$ taken at different temperatures exhibit the same absorption. Such a point implies the existence of two spectral components which are in equilibrium at a definite frequency (energy) of the spectrum. Moreover this means that each species contributes independently and the overall spectrum can be decomposed exactly into the sum of the two fractions. Such a system will exhibit also a Van't Hoff behaviour, i.e. the constant of equilibrium of the reaction (here the equilibrium between the two components of water) is related to the Gibbs energy variation ΔG at the equilibrium by $\Delta G_{equil} = -RT \ln K_{equil}$. The Van't Hoff plot can be drawn by bisecting the spectrum at the isosbestic point and plotting the natural logarithm of the ratio of the areas above and below that point versus 1/T. In case we assume the area below (lower wavelength) the isosbestic point coincident with $I_1(\omega,T)$ and the area above (higher wavelength) coincident with $I_2(\omega,T)$, we obtain a straight line whose slope corresponds to the energy difference between the two vibrational states 1 and 2:

$$\ln \frac{I_1(\omega,T)}{I_2(\omega,T)} = \frac{E_2 - E_1}{RT} + c \qquad (2)$$

Fig. 3 shows the Van't Hoff plot for the model of the two populations, namely interacting and non interacting (including the intermediate population) in a range 298-333 K°C. The picture shows a perfect linearity thus confirming that the difference of energy of the two phases does not depend on temperature in the range under investigation. The slope of the curve ΔE/R (R is the ideal gas constant) gives $\Delta E = 2.9 \pm 0.6 KCal/mol$ which is in reasonable agreement with various estimates in literature (see in particular Table II in ref.8); this value is usually associated to the enthalpy change in the hydrogen-bond rupture for water, however, this value is quite lower than the value 4.5 KCal/mol estimated from the enthalpy of sublimation of ice. The intercept of the straight line is equal to the molar entropy change ΔS/R between the two populations, here $T\Delta S = 2.7 \pm 0.2 KCal/mol$ per particle at T= 298 K, which almost coincides with the above estimate for ΔE showing that the decrease of energy of the fraction (1) with respect to the fraction (2) comes almost exclusively from the vanishing of the entropic component.

Several authors [1,2] have interpreted the existence of the isosbestic point as the result of changes in a continuous distribution of H bonding geometries and of the microscopic nature of the fluctuations of the (single state) thermal bath. They also stated that a single specie coupled to a Gaussian bath satisfies the Van't Hoff behaviour equation. However the demonstration always implies the existence of two species artificially dividing the spectrum at a cut-off frequency. In order to guarantee that the energy difference in eq (2) depends weakly on the temperature, it must be required that the fluctuations are dominated by a few bath modes. In such a way the origin of the effect is moved from the two-state picture of the liquid structure to the two-state fluctuations of the thermal bath whose energy are below and above the average.

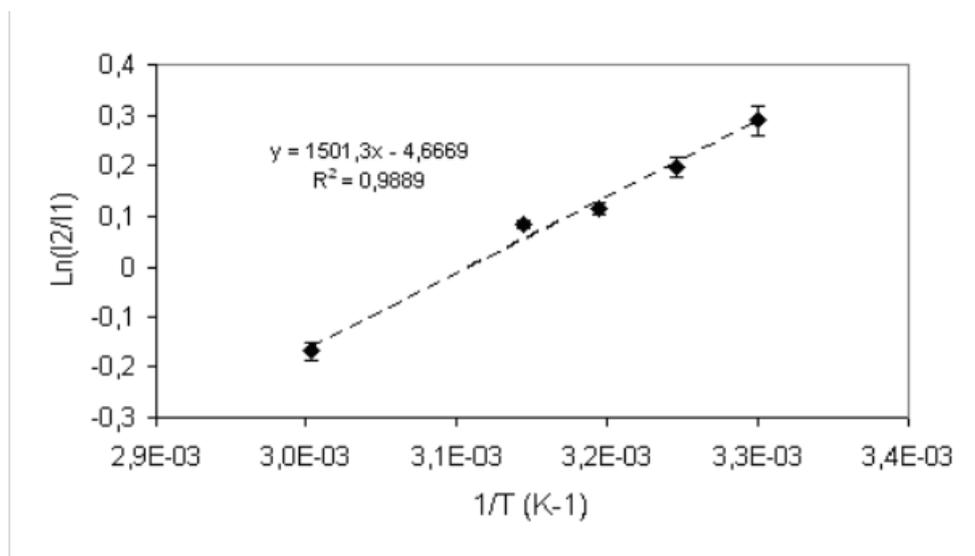

Figure 3. Van't Hoff plot in the range 298 to 333 K.

The analysis of the three populations described above can be used to shed some light on the structuring of water close to hydrophilic surfaces. We have investigated the case of an interesting variety of liquid water reported in literature, namely the case of the so called Exclusion Zone (EZ) water examined by the group led by G.H. Pollack [16,17]. This water is detectable close to hydrophilic surfaces, such as Nafion, and exhibits properties quite different from those found in normal bulk water: a) EZ water is unable to host solutes, and this is the root of the name Exclusion Zone; b) its viscosity is much higher than that of normal water suggesting the presence of a strong interaction among molecules; c) it is an electron-donor, a chemical reducer, whereas normal water is a mild oxidant; consequently the interface EZ water/normal water is a red-ox pile, where the red-ox potential could have a jump of a fraction of a Volt; d) it exhibits a fluorescent response in the UV region at 270 nm. EZ water should therefore imply a major reorganization of the molecular structure of water, in particular (see property d) with respect to the electronic structure. The observed depths of the layers of EZ water on the surface could reach values as high as 500 μm. We have suspended in normal bi- distilled water small pieces of Nafion (which have been previously rinsed many times to get rid of possible impurities). They produce around them shells of EZ water. Then the FT-IR spectrum of the hydrated Nafion, extracted from the bulk water, has been measured in transmission and compared with the spectrum of bulk water in Fig. 4. In Table 1, where we summarize the results, we show the percentage of the peak areas obtained by curve fitting analysis. It is evident that the vibrational spectrum of water adsorbed on Nafion films is very different from the spectrum of bulk water.

TABLE I  Fractions of the three populations in normal and EZ water

|  | Peak position (±4 cm$^{-1}$) | % |
|---|---|---|
| Bulk water 1 | 3205 | 58 |
| Bulk water interm. | 3361 | 37 |
| Bulk water 2 | 3529 | 4.6 |
| EZ water 1 | 3292 | 76.5 |
| EZ water interm. | 3494 | 4 |
| EZ water 2 | 3610 | 19.7 |

The most striking feature is the decrease of the intermediate fraction from 37% to 4%. Correspondingly there is an increase of the other two fractions which share almost equally the components lost by the intermediate fraction. It is worthwhile to note that the ZnSe crystal, normally used as support in ATR measurement as in Fig. 4a), is a non-hygroscopic material, hence it can be envisaged an effect on the partitioning of the three structures opposite to that of the Nafion, i.e. an increase of the interacting fraction with respect to bulk water. These considerations persuade us of the role of the wall on the restructuring of water.

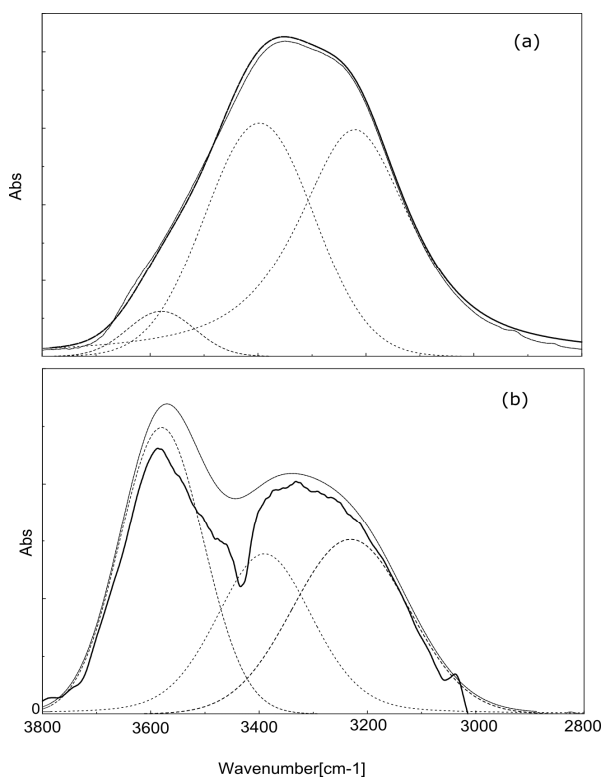

Figure 4. FT-IR peak of the OH stretching : a) liquid water ATR spectrum on ZnSe crystal at room temperature; b) transmission spectrum of wet (dipped in liquid bi-distilled water) Nafion, the dry Nafion spectrum has been used as background. The de-convolution of the spectra has been done using the Jasco curve-fitting software with 15 iterations.

## 3 Discussion

The existence of two different populations within water, made up of molecules having different mutual interactions has been predicted in the framework of Quantum Electrodynamics (QED). It has been shown, starting from first principles, that the interaction among particles is mediated by electromagnetic fields emerging from their collective behaviour [18,19,20,21,22,23]. These fields induce long-range forces which are responsible for the existence of the condensed state. The short-range static attractions come into play only after the molecules have been brought sufficiently close together by the attraction of the long-range force. Since the beginning of the last century it has been accepted that the attractive forces acting between two neutral atoms or molecules separated by a distance large in comparison with their dimensions (Van der Waals Forces) are electromagnetic in origin. A theory of long-range attractive forces in condensed matter was first proposed in 1961 by Dzyaloshinskii, Lifshitz and Pitaevskii, using Quantum Field [24]. They proposed that "the interaction between bodies is considered to take place through a fluctuating electromagnetic field". However this interaction does not induce a phase correlation among components. Actually, many years before, in 1916, Walther Nernst had suggested the possibility of tuning the fluctuations of all the components of a system together in order to produce the appearance of a common phase [25]. Several decades later, using QED has been shown [21] that fluctuating electromagnetic fields not only mediate the interactions among non polar molecules, but, above a density threshold and below a critical temperature, also induce the appearance of a coherent state of the matter in which particles oscillate coherently between two quantum states giving rise to a long-range attraction among atoms or molecules which adds up to their electrostatic properties. In condensed matter therefore two phases exist. One phase is made up of Coherence Domains (CD) in which molecules oscillate coherently in phase with a self trapped em field. The size of each CD is the wavelength of the em mode responsible for the oscillation. The energy per particle in the CDs is lower than the energy of a non coherent particle by the energy gap $E_g$. The other phase is made up by molecules set out of tune by thermal fluctuations. At temperatures other than T= 0, electrodynamic interaction is counteracted by thermal collisions, which push components out of the coherent state. The interplay between the electrodynamic attraction and thermal disruption produces a continuous crossover of molecules between the two regimes A dynamical distribution between the two phases $F_c$, $F_{nc}$ of coherent and non-coherent molecules is thereby established, depending on temperature:

$$F_c(T) + F_{nc}(T) = 1 \qquad (3)$$

According to this view, each atom or molecule belongs to one of the two fractions in a dynamical sense, i.e. it fluctuates between a coherent and a non-coherent state with a characteristic time life $\hbar/E_g$. The particular case of water deserves special consideration. The shape of $F_c(T)$ has been calculated starting from the statistical distribution of the molecules and from the energetic gap $E_g$ [18],[20] and can be now compared with the experimental results. In Fig. 5 the fraction of coherent water calculates as the ratio of height of the peak centered at (about) 3200 cm$^{-1}$ on the sum of the three peaks, is reported versus T in the range 298-333 K and compared to the calculated $F_c(T)$ showing a satisfactory agreement.

As shown in refs. 23-25, molecules in coherence domains are involved in a coherent oscillation between the molecular ground state and an excited state at E= 12.07 eV which lies just below the ionisation threshold (E=12.60 eV). It has also been shown that the properties of such coherent phase are compatible with the phenomenological properties of liquid water; in particular they are able to account for the well known fact that water in living organisms is a reducer which means that it releases electrons [26]. Even the reducing property of the EZ zone can be understood by identifying it with a coherent region. These properties cannot be understood in a picture where molecules are assumed to be non correlated and subjected to short-range interactions only. Moreover, the calculations show [22] that the energy gap of the coherent water at T=0 is: $\Delta E = 0.17 \pm 0.05 eV$. This value must be compared to the energy gap measured by the Van't Hoff plot between interacting and non interacting fraction of dipoles in the FT-IR spectra, $\Delta E$= 2.9 KCal/mol which corresponds to 0.12 eV per molecule.

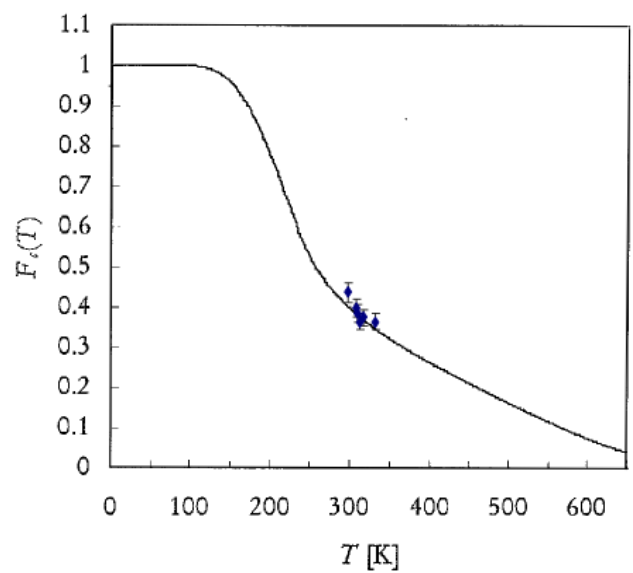

Figure 5. Comparison between the experimental and calcolated coherent fraction versus T [20]. This curve also fits the value $F_c$, $F_{nc}$(T=300K) exposed in ref. 22.

The above fair agreement allows us to identify the fraction (1) with the coherent fraction of water and the fraction (2) with the non-coherent fraction. It must be noted that the value 0.17 eV has been calculated at T= 0 K where the boundaries of CD are sharp and the cross-over is not allowed by thermal fluctuations. At T≠ 0 actually, the boundaries are smeared out by thermal collisions and the energy gap is decreased. In such a picture, even the so called intermediate population could find a rationale. The measured spectrum emerges from a dipole-dipole transition between two specific quantum states. We have seen that the first and third peak describe the transitions occurring within the coherent and non coherent fraction respectively. It is very natural to assign the second (intermediate) peak to the transitions where the initial state is in the coherent fraction and the final state is in the non-coherent fraction and vice versa. In fact, the average life time of the coherent state is $\sim 4 \cdot 10^{-15}$ sec which is about 2 times the vibration transition time scale. Such an interpretation tells us that in the EZ water the two fractions get stabilized since the attraction with walls shields the coherent fraction from the thermal collisions by deepening the energy gap. Another topic to be addressed is the evaluation of the shifts of frequency between the three peaks. The frequency ν of an harmonic oscillator is given by: $2\pi\nu = \sqrt{k/m}$, where $k$ is the spring constant and $m$ is the proton mass. The strength of the spring constant is given by the steepness of the potential well of the proton and is determined by the electron charge distribution in the proton neighbourhoods. The measured O-H distance in the vapour state is $r_{meas}$=0.957 [27], whereas in the liquid phase it is $r_{meas}$ =1.01Å [28]. By assuming that the liquid phase is composed by a mixture of a coherent fraction $F_c$ and a non-coherent fraction $F_{nc}$ according to (3) the following relation holds :

$$\langle r_{measured} \rangle = F_{nc} \langle r_{nc} \rangle + F_c \langle r_c \rangle \quad (4)$$

Arani et al. calculated that at room temperature $F_c$= 0.424 and $F_{nc}$= 0.576 [20], assuming that the O-H distance in the non coherent state equals $r_{meas}$ in the vapour phase and substituing in the (4) we can estimate the O-H distance in the coherent phase as: $\langle r_c \rangle = 1.082$ Å.

Please note that the equilibrium position of the proton is different in the coherent and non-coherent phases. However this difference, $\Delta l = r_c - r_{nc} = 0.125$ Å is smaller than the uncertainty of the proton position, given by $\sqrt{\hbar/m\omega}$ Å (assuming an oscillation frequency ω of about 3600 cm$^{-1}$). We can now estimate the variation of the spring constant from the elongation $l$ of the bond from the relations:

$$\frac{\Delta K}{K} = -\frac{\Delta l}{l_{nc}} \approx -\frac{0.125}{0.975} \approx -0.130 \qquad (5)$$

we find the variation of the oscillation frequency: $\frac{\Delta\omega}{\omega} = -\frac{1}{2}\frac{\Delta l}{l} = -0.065$ which is quite close to the experimental value $\frac{\Delta\omega}{\omega} = \frac{\omega_{nc} - \omega_c}{\omega_{nc}} = 0.092$ (frequencies are from Table 1). The position of the intermediate peak halfway between the two other peaks can be understood as the IR emission occurring during the transition of the water molecule from the coherent fraction to the non-coherent fraction and vice-versa. The reliability of the above argument is shown also by the consideration that the bending transition occurring at 1650 cm$^{-1}$ does not exhibit the same broadening as the stretching transition. Actually the bending motion does not imply any variation of the bonding length.

It can also be observed that the position of the coherent peak almost doesn't change with temperature, whereas the position of the non coherent peak is subjected to significant changes. As a matter of fact, the internal dynamics of coherent fraction, to which molecules of the lowest peak belong, does not depend on T because the energy gap protects molecules from thermal collisions (in fact $E_{gap}$ ~6.8 KT), contrary to the non coherent dynamics. Finally the small shift of the absolute values of the peak positions in case of contact with an hydrophilic surface, as observable in Table 1, may be explained by the different strength of the interaction of water molecules with the wall for coherent and non coherent molecules.

4 Conclusions

The present analysis of the vibrational spectra of liquid water shows a remarkable agreement with the predictions of QED theory. For the first time it has been shown, without introducing ad-hoc parameters, that experimentally measured physical variables such as the energy gap between the two populations of molecules and the broadening of the stretching peak coincide with the values calculated in the frame of electrodynamics of liquid water.

The structure of liquid water proposed either by MD modelling of the shapes of IR spectra or by the decomposition of the total population in two independent ensembles of molecules having different enthalpies, lies on the fundamental assumption of an ensemble of isolated chromophores. The picture of liquid water emerging from QED is, at the opposite, based on the existence of long-range correlations among molecules due to their quantum nature.

A very debated issue, the origin of the isosbestic point, has been attributed to a discrimination of frequency below and above a certain cut-off frequency which undergoes to the requirement to be weakly dependent on temperature [4]. In the present paper we are able to explain the above requirement dynamically: the cut-off frequency coincides mathematically with the temperature independent energy gap provided by the coherence equations in [19]. The existence of such a gap, which also coincides with the energy difference between the two populations in equilibrium at the isosbestic point, cannot be obtained with a continuous distribution of H-bonds related to each type of molecules.